\documentclass[preprint,prd,noshowpacs,nofootinbib]{revtex4}

\usepackage{color} 
\usepackage{amssymb}
\usepackage{amsmath}
\usepackage{graphicx} 
\usepackage{float}
\usepackage{braket}     
\usepackage{pdfpages}
\usepackage{epstopdf}
\usepackage{soul}
\usepackage[utf8]{inputenc}

\begin{document}

\title{Symmetrized operators or modified integration measure in Generalized Uncertainty Principle Models}

\author{Michael Bishop}
\email{mibishop@mail.fresnostate.edu}
\affiliation{Mathematics Department, California State University Fresno, Fresno, CA 93740}

\author{Daniel Hooker}
\email{dshooker@mail.fresnostate.edu}
\affiliation{Physics Department, California State University Fresno, Fresno, CA 93740}

\author{Douglas Singleton}
\email{dougs@mail.fresnostate.edu}
\affiliation{Physics Department, California State University Fresno, Fresno, CA 93740 }

\date{\today}

\begin{abstract}
Many Generalized Uncertainty Principle (GUP) models modify the inner-product measure to ensure symmetric position or momentum operators. We show that an alternate approach to these GUPs is to symmetrize the operators rather than modifying the inner product. This preserves the standard momentum space allowing the eigenstates and maximally localized states of the modified position operator to have a standard position representation. We compare both approaches and highlight their merits.
\end{abstract}

\maketitle

\section{Introduction}

The Generalized Uncertainty Principle (GUP) is a phenomenological approach to quantum gravity \cite{amati,maggiore,kmm,garay,scardigli,adler-1999,adler-2001} that modifies the standard commutator, $[{\hat x}, {\hat p}] = i\hbar$, to incorporate gravitational effects. Such modifications often imply a minimum length, consistent with predictions from string theory and loop quantum gravity \cite{string,loop}.
A well-known deformation of the standard commutator from Kempf, Mangano, and Mann (KMM) \cite{kmm} is
\begin{eqnarray}
    \label{comm-deform}
    [{\hat X}, {\hat P}] = i \hbar (1 + \beta \hat{p}^2) = i \hbar (1 + \beta p^2) ~, 
\end{eqnarray}
where in the last expression we have written the position and momentum operators using ${\hat x} = i \hbar \frac{d}{dp}$ and ${\hat p}=p$.
The parameter $\beta$ has dimensions of inverse momentum squared and sets the scale for gravitational corrections to the commutator. Typically, one expects $\beta$ to be of the order of $1/({\rm Planck~ momentum})^2$. One can also take a phenomenological approach and leave $\beta$ as a parameter to be probed by experiment or observation. An up-to-date review of the experimental/observational constraints on $\beta$ can be found in \cite{bosso1}.

The commutator in \eqref{comm-deform} requires modifying the position operator, the momentum operator, or both. There is some debate in the literature around this issue.  Reference \cite{BLS} notes that there are numerous ways to modify the position and/or momentum operators to obtain the same modified commutators and argues that these theories often have different physical consequences. A contrasting view can be found in reference \cite{bosso2}, which argues that the physical consequences mentioned above are instead independent of the form of the modified commutators. 
In this paper, we focus on the mathematical and theoretical differences between the choice of modified operators from \cite{kmm} and the symmetrized version of these operators, which are given below in equation \eqref{x-opertor}.
The form of the modified operators originally proposed in \cite{kmm} that gave \eqref{comm-deform} was
\begin{equation}
\label{kmm-ori}
{\hat X}_{KMM} = (1+\beta {\hat p}^2)\hat{x} = i \hbar  \left( 1 + \beta  p^2 \right) \frac{d}{dp} ~~~{\rm  and}~~~ {\hat P} = {\hat p} = p.
\end{equation}
The operator $\hat{X}_{KMM}$ is neither symmetric nor Hermitian, i.e., $\hat{X}_{KMM}^\dagger \neq \hat{X}_{KMM}$.
In contrast to \eqref{kmm-ori}, we define the operators as 
\begin{equation}
\label{x-opertor}
    {\hat X}_{sym} =  \hat{x} + \beta  \hat{p}\hat{x}\hat{p}  ~~~{\rm and}~~~ {\hat P} = {\hat p} = p  ~,
\end{equation}
where we capitalize the modified position and momentum operators and leave the standard quantum operators in lower case.
Substituting the operators from \eqref{x-opertor} into $[{\hat X}, {\hat P}]$ reproduces the commutator in \eqref{comm-deform}.

The generalized commutator from \eqref{comm-deform} and the modified operators in \eqref{x-opertor} lead to the same modified uncertainty principle as in \cite{kmm}, namely
\begin{equation}
\label{p-GUP}
    \Delta X \Delta p \ge \frac{\hbar}{2} (1 + \beta \langle p ^2 \rangle ) \to \frac{\hbar}{2} (1 + \beta \Delta p^2).
\end{equation}
In the last step, we substituted $\langle p ^2 \rangle = \Delta p ^2  + \langle p \rangle ^2$ and found a lower bound by taking $\langle p \rangle =0$. Solving \eqref{p-GUP} for $\Delta X$ gives $\Delta X \ge \frac{\hbar}{2} \left( \frac{1}{\Delta p} + \beta   \Delta p\right)$. This relationship implies a minimum length $\Delta X_{min} = \hbar \sqrt{\beta}$ that occurs at $\Delta p = \frac{1}{\sqrt{\beta}}$.  Although the operators in \eqref{x-opertor} differ from those of KMM \eqref{kmm-ori}, they lead to the same commutator, the same generalized uncertainty principle, and the same minimal length.

\section{Fundamental Difference between Modified Inner Product and Symmetric Modified Operator}

In KMM \cite{kmm}, the modified position operator $\hat{X}_{KMM}$ is not symmetric under the standard inner product ({\it i.e.}, $\bra{\psi} ({\hat X}_{KMM} \ket{\phi} ) \ne (\bra{\psi \hat X}_{KMM} ) \ket{\phi}$), thus it may have complex eigenvalues. 
To address this issue, reference \cite{kmm} proposed modifying the inner product:
\begin{equation}
    \label{inner1}
    \bra{\psi} \phi \rangle _{KMM} = \int_{-\infty} ^\infty  \frac{\psi^{\dagger} (p) \phi (p) }{1 + \beta p^2} dp ~.
\end{equation}
The symmetry of the momentum operator from \eqref{x-opertor} ({\it i.e.}, $\bra{\psi} ({\hat P} \ket{\phi} ) = (\bra{\psi \hat P} ) \ket{\phi}$) works with or without the modified measure in \eqref{inner1} since ${\hat P} = {\hat p}=p$. However, the symmetry of the modified position operator, ${\hat X}_{KMM} = i \hbar  \left( 1 + \beta  p^2 \right) \frac{d}{dp}$ requires the modified measure in order for $\bra{\psi} ({\hat X}_{KMM} \ket{\phi} _{KMM}) = (\bra{\psi \hat X}_{KMM} ) \phi \rangle_{KMM}$ since the factor of $ \left( 1 + \beta  p^2 \right)^{-1}$ in \eqref{inner1} is required to cancel the factor $\left( 1 + \beta  p^2 \right)$ in the modified position operator, which would otherwise lead to additional terms when doing the integration by parts. 

The modified measure in \eqref{inner1} breaks the isometry between the momentum-space and position-space wavefunctions. The modified inner product redefines the momentum space into something mathematically distinct from the standard momentum space. In conventional quantum mechanics, the Fourier transform provides a unitary map between position-space wavefunctions $\Psi(x)$ and their momentum-space counterparts $\psi(p)$:
\begin{equation}
    \label{mom-pos}
    \Psi (x) = \frac{1}{\sqrt{2 \pi \hbar}} \int_{-\infty}^\infty \psi (p) e^{i p x /\hbar} d p ~~~~ \longleftrightarrow ~~~~ \psi (p) = \frac{1}{\sqrt{2 \pi \hbar}} \int_{-\infty}^\infty \Psi (x) e^{-i p x /\hbar} d x
\end{equation}
where uppercase corresponds to position space and lowercase corresponds to momentum space.  
With the new inner product \eqref{inner1}, the Fourier transform ceases to represent an inner product with plane waves indexed by $x$. Consequently, the unitary correspondence between standard position and modified momentum spaces is lost.

As a result, there is no longer a natural representation for GUPs in position space.
The authors of \cite{kmm} were well aware of this issue and developed an alternative to the Fourier transform by defining `maximally-localized states' and `quasi-position wavefunctions.'  The new transforms (equations (43) and (47) in \cite{kmm} or equations (9a) and (9b) in \cite{bosso}) define maps between the modified momentum space and `quasi-position' space.  
The major issue is that there is no method for mapping the quasi-position space into standard position space.  
A detailed analysis \cite{bosso} of quasi-position space in GUPs showed that any approach to GUP that modifies the inner product does not have a standard position representation.
Some alternative operator modifications aim to restore a position-space representation without altering the momentum-space inner product \cite{BLS}. For instance, defining $\hat{X}=\hat{x}=i\hbar \frac{d}{dp}$ and $\hat{P} = p + \frac{\beta}{3}p^3$ avoids redefining the measure; however, this system does not have a minimal length, which is the central motivation for GUPs.  

In contrast, by symmetrizing ${\hat X}$ as in \eqref{x-opertor}, there is no need to modify the inner product of momentum space to ensure a real spectrum.  This choice preserves the Fourier transform, which means that any state in momentum space has a physically and unitarily equivalent state in position space. 
There is no longer a need for the `quasi-position' spaces.  Instead, the eigenfunctions of the modified position operator (or the maximally-localized states) can be used to form a basis for the standard $L^2$ position and momentum spaces as we will demonstrate below.  These functions have both standard position and momentum representations.  This approach resolves the difficulty in representing GUPs in position space.  

\section{Eigenfunctions and Maximally Localized States for the Symmetric Position Operator}

We now obtain the eigenstates and maximally localized states for the symmetrized position operator and compare them with the corresponding versions in KMM \cite{kmm}.
  
\subsection{Modified position eigenstates}
The eigenfunctions of the symmetric position operator satisfy
\begin{equation}
    \label{eigen2}
    \hat{X}_{sym}\psi_{\xi}(p) 
    = i\hbar\frac{d}{dp}[\psi_\xi(p)] + \beta p \frac{d}{dp}[p\psi_{\xi}(p)] 
    = \xi\,\psi_{\xi}.
\end{equation}
The solutions are
\begin{equation}
    \label{soln2}
    \psi_{\xi}(p) = 
    \sqrt{\tfrac{\sqrt{\beta}}{\pi}}
    \frac{1}{\sqrt{1+\beta p^2}}
    \exp\!\left[-\,\tfrac{i\xi}{\hbar\sqrt{\beta}} 
    \arctan(\sqrt{\beta}\,p)\right],
\end{equation}
with normalization factor $\sqrt{\sqrt{\beta}/\pi}$. For comparison, the KMM eigenstates for $\hat{X}_{KMM}$ are $\psi^{KMM}_{\xi}(p) = \sqrt{\frac{\sqrt{\beta}}{\pi}}   \exp\!\left[-\,\tfrac{i\xi}{\hbar\sqrt{\beta}}  \arctan(\sqrt{\beta}\,p)\right]$ \cite{kmm}. 
These position eigenstates for $\hat{X}_{KMM}$ lack the $1/\sqrt{1+\beta p^2}$ factor in the denominator of the eigenstates of the symmetric position operator. In KMM, this factor reappears through the modified inner product, ensuring $\langle \psi_{\xi}|\psi_{\xi'} \rangle_{KMM}
= \langle \psi_{\xi}|\psi_{\xi'} \rangle_{sym}$.  
Using either $\hat{X}_{sym}$ or $\hat{X}_{KMM}$ yields similar calculations, yet their treatment of momentum space is fundamentally different.

The scalar product of two such states is
\begin{eqnarray}
    \label{ortho}
    \langle \psi_{\xi}|\psi_{\xi'} \rangle 
    &=& \frac{\sqrt{\beta}}{\pi} \int_{-\infty}^{\infty}
        \frac{\exp\!\left[-\,\tfrac{i(\xi'-\xi)}{\hbar\sqrt{\beta}}
        \arctan(\sqrt{\beta}\,p)\right]}{1+\beta p^2}dp \nonumber \\
    &=& \frac{2\hbar\sqrt{\beta}}{\pi(\xi'-\xi)}
        \sin\!\left[\tfrac{(\xi'-\xi)\pi}{2\hbar\sqrt{\beta}}\right].
\end{eqnarray}
where the factor of $1+\beta p^2$ in denominator is due to the position eigenfunctions rather than a modified integration measure.
These states are not mutually orthogonal, unlike plane waves in standard quantum mechanics. We can generate orthogonal sets by restricting $\xi$ to a discrete set spaced by $2\hbar\sqrt{\beta}$.  That is, choose any $\epsilon \in [-1,1]$ and define $\xi_n := (2n+\epsilon)\hbar\sqrt{\beta}$ where $n\in\mathbb{Z}$.  Then $\xi_n-\xi_{n'} = 2(n-n')\hbar\sqrt{\beta}$ and
\begin{equation}
    \label{discrete}
    \langle \psi_{\xi_n} | 
    \psi_{\xi_{n'}} \rangle = \delta_{nn'}.
\end{equation}
Thus, the $\xi_n$ serve as a discrete lattice in modified position.  However, each $\xi_n$ lattice point corresponds to an eigenfunction defined on all of $\mathbb{R}$. 

We now show that any $\psi_{\xi}$ lies in the span of the orthogonal set $\{\psi_{2n\hbar\sqrt{\beta}}: n\in\mathbb{Z}\}$. This requires
\begin{equation}
    \label{e1}
    |\psi_{\xi}\rangle 
    \stackrel{?}{=} \sum_{n=-\infty}^{\infty}
    \langle \psi_{2n\hbar\sqrt{\beta}}|\psi_{\xi}\rangle
    |\psi_{2n\hbar\sqrt{\beta}}\rangle.
\end{equation}
Applying $\bra{\psi_\xi}$ to both sides gives
\begin{equation}
\label{e2}
|\psi_{\xi}|^2 =1 \stackrel{?}{=} \sum _{n=-\infty} ^{\infty} | \bra{\psi_{{2n \hbar \sqrt{\beta}}} } \psi_\xi \rangle|^2  ~.
\end{equation}
We can rewrite $\xi=(2n'+\epsilon)\hbar\sqrt{\beta}$ for some $n'\in\mathbb{Z}$ and $-1\leq\epsilon\leq1$, we obtain
\begin{eqnarray}
\label{e3}
    \sum _{n=-\infty} ^{\infty} | \bra{\psi_{{2n \hbar \sqrt{\beta}}} } \psi_\xi \rangle|^2 &=& \sum _{n=-\infty} ^{\infty} \left|\frac{2\hbar\sqrt{\beta}}{\pi(\xi_n-\xi)} \sin\left(\tfrac{\pi(\xi_n - \xi)}{2\hbar\sqrt{\beta}}\right)\right|^2 \nonumber \\
    &=& \sum _{m=-\infty} ^{\infty} \frac{4}{\pi^2}\frac{1}{(2m+\epsilon)^2} \sin^2\left(\tfrac{2m+\epsilon}{2}\pi\right) \nonumber\\
    &=&  1 ~,
\end{eqnarray}
where we shifted $n\to m=n'-n$ and evaluated the sum with {\it Mathematica}.  
Thus, every $\psi_{\xi}$ lies in the span of the orthogonal basis $\psi_{\xi_n}$.

In fact, any of these orthogonal sets is a basis for momentum space $L^2(\mathbb{R})$.  To see this, we first show the family of eigenstates $\{\psi_\xi:\xi\in\mathbb{R}\}$ is complete.  To see this, consider the following unitary map from $L^2(\mathbb{R})$ to $L^2\left(-\frac{\pi}{2},\frac{\pi}{2}\right)$:
\begin{equation}
    \label{comp1}
    U(f)(t) := \beta^{-1/4}\sec(t)\, f\!\left(\tfrac{\tan(t)}{\sqrt{\beta}}\right).
\end{equation}
To verify this map is unitary, for any two functions $f$ and $g$ in $L^2(\mathbb{R})$, we have
\begin{eqnarray}
    \label{comp2}
    \langle U(f), U(g)\rangle_{L^2\left(-\frac{\pi}{2},\frac{\pi}{2}\right)} 
    &=& \int_{-\pi/2}^{\pi/2} \overline{\sec(t) f\!\left(\tfrac{\tan(t)}{\sqrt{\beta}}\right)} 
    \sec(t) g\!\left(\tfrac{\tan(t)}{\sqrt{\beta}}\right) \beta^{-1/2}\, dt \nonumber\\
    &=& \int_{-\infty}^\infty \overline{f(p)}\, g(p)\, dp \\
    &=& \langle f, g\rangle_{L^2(\mathbb{R})}, \nonumber
\end{eqnarray}
where we used the substitution $p=\tan(t)/\sqrt{\beta}$ for the integral from the first to the second line.  This unitary map sends the eigenfunctions to the plane waves:
\begin{eqnarray}
    \label{comp3}
    U(\psi_\xi)(t) 
    =\beta^{-1/4}\sec(t)\sqrt{\tfrac{\sqrt{\beta}}{\pi}} 
    \frac{\exp\!\left[-i\tfrac{\xi}{\hbar\sqrt{\beta}}
    \arctan\!\big(\tfrac{\sqrt{\beta}\tan(t)}{\sqrt{\beta}}\big)\right]}
    {\sqrt{1+\beta \left(\tfrac{\tan(t)}{\sqrt{\beta}}\right)^2}} = \frac{\exp\!\left[-i\tfrac{\xi}{\hbar\sqrt{\beta}}t\right]}{\sqrt{\pi}}.
\end{eqnarray}
Because the plane waves are complete in $L^2\left(-\frac{\pi}{2},\frac{\pi}{2}\right)$, for any function $f(p) \in L^2(\mathbb{R})$, $U(f)(t)$ can be decomposed into an integral decomposition of plane waves.  Applying $U^{-1}$ to the integral decomposition of $U(f)$ gives the integral decomposition of $f$ in terms of the eigenfunctions $\psi_\xi$.  Therefore, the eigenfunctions form a complete set. 
Moreover, the restriction to the eigenfunctions indexed by $\xi_n$  (with $\epsilon =0$) corresponds to the set of plane waves with integer wavelength in $L^2\left(-\frac{\pi}{2},\frac{\pi}{2}\right)$.  Since the countable set of eigenfunctions $\psi_{\xi_n}$ span the entire set of eigenfunctions which are complete, the countable set also spans all of $L^2(\mathbb{R})$ and forms an orthonormal basis.

Finally, since the symmetric position operator in \eqref{eigen2} do not require a change in the inner product as in \eqref{inner1} one can obtain the position space version of the momentum space eigenfunctions in \eqref{soln2} using the usual Fourier transformation between position and momentum space given in \eqref{mom-pos}. Doing this gives the position space representation of \eqref{soln2} as:
\begin{align}
    \label{FT1}
    \Psi_{\xi}(x) 
    &= \frac{1}{\sqrt{2\pi\hbar}}
       \int_{-\infty}^{\infty}
       e^{ipx/\hbar}\, \psi_{\xi} (p) \,dp \nonumber \\
       &= \frac{1}{\sqrt{2\pi\hbar}} \sqrt{\frac{\sqrt{\beta}}{\pi}}
       \int_{-\infty}^{\infty}
        \frac{e^{ipx/\hbar}}{\sqrt{1+\beta p^2}}
    e^{\!-\,i\xi \left[\arctan(\sqrt{\beta}\,p)/\hbar\sqrt{\beta} 
    \right]}\,dp~ \\
    &= \frac{\sqrt{2 \sqrt{\beta}}}{ \pi \sqrt{\hbar} } K_{\frac{i |\xi|}{\hbar \sqrt{\beta}}} \left( \frac{|x|}{\hbar \sqrt{\beta}}\right) \nonumber
\end{align}
where $K_{\frac{i |\xi|}{\hbar \sqrt{\beta}}}$ is a modified Bessel function of the second kind of order $\frac{i |\xi|}{\hbar \sqrt{\beta}}$. Taking $\xi=0$, the final result in \eqref{FT1} becomes
\begin{align}
\label{FT3}
     \Psi_0(x) =  \frac{\sqrt{2 \sqrt{\beta}} }{\pi\sqrt{\hbar}} K_0 \left( \frac{|x|}{\hbar \sqrt{\beta}} \right)~.
\end{align}
Taking the limit $\beta \to 0$ one can show that $\Psi_0 (x) \sim \delta (|x|)$, {\it i.e.} $\Psi_0 (x)$ becomes a Dirac delta which is the expected result in the limit of ordinary quantum mechanics. 

\subsection{Maximally Localized States}

Maximally localized states are defined as those centered at position $\xi$ with minimum uncertainty $\Delta X_{min}$:
\begin{equation}
    \label{MLa}
    \langle \psi^{ML}_\xi |X| \psi^{ML}_\xi \rangle = \xi, 
    \qquad 
    \Delta X_{\psi^{ML}_\xi} = \Delta X_{min} = \hbar \sqrt{\beta}.
\end{equation}
From \cite{kmm}, maximally localized states satisfy
\begin{equation}
    \label{33}
    \Big[ \hat{X} - \langle X \rangle 
    + i\hbar \frac{1 + \beta \Delta p^2 + \beta \langle p \rangle}{2\Delta p^2}(p-\langle p \rangle)\Big] 
    \psi^{ML}_\xi(p) = 0.
\end{equation}
The minimal length arises for $\langle X\rangle=\xi$, $\langle p\rangle=0$, and $\Delta p = 1/\sqrt{\beta}$.  
Using $\hat{X}\to\hat{X}_{sym}$ gives
\begin{equation}
    \label{ML3}
    \psi^{ML}_\xi(p) 
    = \sqrt{\frac{2\sqrt{\beta}}{\pi}}
      \frac{1}{1+\beta p^2}
      \exp\!\left[-\,\tfrac{i\xi}{\hbar\sqrt{\beta}}
      \arctan(\sqrt{\beta}p)\right],
\end{equation}
with normalization factor $\sqrt{2\sqrt{\beta}/\pi}$.

These states differ from those in \cite{kmm} only in the prefactor, $1/(1+\beta p^2)$ instead of $1/\sqrt{1+\beta p^2}$. The normalization integrals are otherwise identical. As with the position eigenstates $\psi_{\xi}(p)$, the maximally localized states are not orthogonal. From \eqref{ML3}:
\begin{eqnarray}
    \label{ML4}
    \bra{\psi^{ML}_{\xi} }\psi^{ML}_{\xi'} \rangle &=& \int _{-\infty} ^\infty dp (\psi^{ML}_{\xi}) ^{\dagger}(p) \psi^{ML}_{\xi'}(p)  \nonumber \\
    &=&\frac{2 \sqrt{\beta}}{\pi} \int _{-\infty} ^\infty  \frac{dp}{(1 + \beta p^2)^2} \exp\left[- i \tfrac{(\xi' - \xi)}{\hbar \sqrt{\beta}} \arctan (\sqrt{\beta} p) \right]  \\
    &=& \frac{1}{\pi} \left[ \tfrac{(\xi ' - \xi)}{2 \hbar \sqrt{\beta}} -  \left(\tfrac{(\xi ' - \xi)}{2 \hbar \sqrt{\beta}}\right)^3 \right]^{-1} \sin \left( \tfrac{(\xi ' - \xi) \pi}{2 \hbar \sqrt{\beta}}\right) \nonumber ~.
\end{eqnarray}
The result in \eqref{ML4} matches \cite{kmm} (eqs.~39–41), except here the factor $(1+\beta p^2)^{-2}$ arises entirely from $\psi^{ML}_\xi$, rather than partly from the measure. Like $\psi_\xi$, the states $\psi^{ML}_\xi$ become orthogonal when $\xi$ is discretized as $\xi_n=2n\hbar\sqrt{\beta}$:
\[
\langle \psi^{ML}_{2n\hbar\sqrt{\beta}}|
\psi^{ML}_{2n'\hbar\sqrt{\beta}}\rangle=\delta_{nn'}.
\]

To show that maximally localized states lie in the span of the discretized position eigenstates, consider
\begin{eqnarray}
    \label{MLspan1}
    \langle \psi_{\xi'}|\psi^{ML}_\xi\rangle
    &=& \frac{\sqrt{2\beta}}{\pi}\int_{-\infty}^\infty
        \frac{dp}{(1+\beta p^2)^{3/2}}
        \exp\!\left[-\,\tfrac{i(\xi-\xi')}{\hbar\sqrt{\beta}}
        \arctan(\sqrt{\beta}p)\right] \nonumber \\
    &=& \frac{2\sqrt{2}\cos\!\big(\tfrac{(\xi-\xi')\pi}{2\hbar\sqrt{\beta}}\big)}
            {\pi\!\left[1-\big(\tfrac{\xi-\xi'}{\hbar\sqrt{\beta}}\big)^2\right]}.
\end{eqnarray}
With discretization $\xi'=2n'\hbar\sqrt{\beta}$, $\xi=2n\hbar\sqrt{\beta}$, this reduces to
\begin{equation}
    \label{MLspan2}
    \langle \psi_{2n'\hbar\sqrt{\beta}}|
    \psi^{ML}_{2n\hbar\sqrt{\beta}}\rangle
    = \frac{2\sqrt{2}\cos(m\pi)}{\pi(4m^2-1)},
\end{equation}
where $m=n-n'$.
Summing the squared overlaps yields
\[
\sum_{n=-\infty}^\infty
\big|\langle\psi_{2n'\hbar\sqrt{\beta}}|\psi^{ML}_{2n\hbar\sqrt{\beta}}\rangle\big|^2
= \sum_{m=-\infty}^\infty
  \frac{8}{\pi^2}\frac{\cos^2(m\pi)}{(4m^2-1)^2}=1,
\]
so the maximally localized states lie in the span of the modified position eigenstates. Thus, no distinct ``quasi-position'' space is required.

Using the Fourier transform, the maximally localized states in position space are given by
\begin{align}
    \label{FT2}
    \Psi^{ML}_\xi(x)
    &= \frac{1}{\sqrt{2\pi\hbar}}
       \int_{-\infty}^{\infty}
       e^{ipx/\hbar}\, \psi^{ML} _{\xi} (p) \,dp  \\
       &= \frac{1}{\sqrt{2\pi\hbar}} \sqrt{\tfrac{2 \sqrt{\beta}}{\pi}}
       \int_{-\infty}^{\infty}
        \frac{e^{ipx/\hbar}}{1+\beta p^2}
    e^{\!-\,i\xi \left[\arctan(\sqrt{\beta}\,p)/\hbar\sqrt{\beta} 
    \right]}\,dp~ \nonumber
\end{align}
Unlike the case of the position eigenstates in \eqref{FT1} we were unable to find a closed form expression for $\Psi^{ML}_\xi(x)$. However, for the special case $\xi =0$ we can obtain a closed form for $\Psi^{ML}_0(x)$ given by
\begin{align}
    \label{FT4}
   \Psi^{ML}_0(x) &=  \frac{1}{\sqrt{2 \pi \hbar}} \sqrt{\tfrac{2 \sqrt{\beta}}{\pi}} \int _{-\infty} ^\infty \frac{e^{ip x / \hbar }}{1+ \beta  p^2} dp = \frac{1}{\beta^{1/4}\sqrt{\hbar}}
       \exp\!\left[-\,\tfrac{|x|}{\hbar\sqrt{\beta}}\right],
\end{align}
In the limit $\beta \to 0$ $\Psi^{ML}_0(x) \sim \delta (|x|)$ {\it i.e.} in the limit when the GUP parameter goes to zero, we recover the standard quantum mechanical result as expected.

Finally, one can show that the maximally localized states are complete. The proof closely follows the completeness proof of the modified position eigenfunctions of the previous subsection; thus, we do not include it here for brevity.

\section{Conclusions}

We compared two formulations of the generalized uncertainty principle (GUP): the modified position operator $\hat{X}_{KMM}$ from \cite{kmm} and its symmetrized version $\hat{X}_{sym}$ defined in \eqref{x-opertor}. Both yield the same modified commutator \eqref{comm-deform}, the uncertainty relation \eqref{p-GUP}, and minimal length. In both cases, the operators satisfy the symmetry conditions $\langle \psi|\hat{X}\phi \rangle = \langle \hat{X}\psi|\phi \rangle$ and $\langle \psi|\hat{P}\phi \rangle = \langle \hat{P}\psi|\phi \rangle$. However, $\hat{X}_{KMM}$ requires a modified inner product \eqref{inner1}, while $\hat{X}_{sym}$ does not. Moreover, $\hat{X}_{sym}$ is Hermitian, whereas $\hat{X}_{KMM}$ is not.

The drawback of modifying the inner product, as in \eqref{inner1}, is the loss of correspondence between standard momentum and position space. The Hilbert space is redefined, and the Fourier transform is no longer a unitary map between the two. Consequently, many GUP models lack a standard position-space representation \cite{bosso}.

In contrast, symmetrizing the operator preserves standard momentum space and thus the usual Fourier transform to position space. The modified position eigenfunctions can be constructed in momentum space and transformed into position space, yielding a consistent representation. By restricting the eigenfunctions to $\xi_n = (2n+\epsilon)\hbar\sqrt{\beta}$, one obtains an orthonormal eigenbasis for the modified position. This effectively discretizes space into a lattice with spacing $2\hbar\sqrt{\beta}$, while standard space remains continuous. \\~\\
{\bf Acknowledgments:} DH and DS acknowledge the support of the Frank Sutton Research Fund.  MB and DS acknowledge support from the Fresno State College of Science and Mathematics RSCA fund.

\end{document}